\documentclass[twocolumn,pre,floatfix]{revtex4}
\usepackage{psfrag,epsfig,amsfonts,amssymb,amsmath,wasysym,bm}
\usepackage{dcolumn}
\usepackage{bbold}
\usepackage[normalem]{ulem}

\usepackage{color}
\usepackage{tabularx}

\usepackage{enumitem} 

\usepackage{hyperref}

\newcommand{\e}{\mathrm{e}}		
\newcommand{\kB}{k_{\mathrm B}} 

\newcommand{\tr}{\mbox{Tr}}

\newcommand{\mic}{\mathrm{mc}}
\newcommand{\rhomic}{\rho_{\mathrm{mc}}}

\newcommand{\rhog}{\rho_{\mathrm{\!\!\;\beta}}}

\providecommand{\norm}[1]{\|#1\|}

\newcommand{\lmat}{\left( \begin{matrix}}	
\newcommand{\rmat}{\end{matrix} \right)}	

\begin{document}

\title{Symmetry-prohibited thermalization after a quantum quench}

\author{Peter Reimann}
\affiliation{Fakult\"at f\"ur Physik, 
Universit\"at Bielefeld, 
33615 Bielefeld, Germany}
\date{\today}

\begin{abstract}
The observable long-time behavior of an isolated many-body system after a 
quantum quench is considered, i.e., an eigenstate 
(or an equilibrium ensemble)
of some pre-quench 
Hamiltonian $H$ serves as initial condition which then evolves in time
according to some post-quench Hamiltonian $H_p$.
Absence of thermalization is analytically demonstrated
for a large class of quite common pre- and post-quench
spin Hamiltonians.
The main requirement is that the pre-quench Hamiltonian must
exhibit a 
$Z_2$ (spin-flip) 
symmetry, which would be spontaneously 
broken in the thermodynamic limit, though we actually 
focus on finite (but large) systems.
On the other hand, the post-quench Hamiltonian 
must 
violate
the $Z_2$ 
symmetry, but for the rest may be non-integrable and may 
obey the eigenstate thermalization hypothesis for (sums of) 
few-body observables.
\end{abstract}

\maketitle

\section{Introduction}
\label{s1}
 
The issue of thermalization in isolated many-body 
quantum systems has been investigated 
for almost a century \cite{neu29,gol10}, leading to 
fascinating
experimental, numerical, and analytical progress 
in recent years \cite{mor18,dal16,gog16,lan16,ued20,nan15}.
The key question in this context is whether or
not the long-time expectation values of experimentally
relevant 
observables are sufficiently well approximated by 
the corresponding microcanonical expectation values, 
as predicted by textbook statistical mechanics.

While 
analytical
results are still rather scarce,
numerical evidence and heuristic arguments
provide quite convincing support for the
common expectation that the problem of 
thermalization is closely connected with 
the issue of integrability and with the 
eigenstate thermalization hypothesis (ETH)
\cite{mor18,dal16,gog16,lan16,ued20,nan15}.
In fact, thermalization, non-integrability, and 
the ETH were 
initially presumed to be 
largely equivalent, but this simple picture
has subsequently been challenged by
various counter-examples, among others
in Refs. \cite{shi17,tur18,lin19,jam19a,jam19b,ban11}, 
though some questions regarding their 
significance are still open 
\cite{mon18,shi18,tur18a,kim15,lin17,far17}.
Further prominent counter-examples are supposed 
to be system exhibiting many-body localization (MBL) 
\cite{mor18,dal16,gog16,nan15},
which in turn have been recently questioned, e.g., 
in Refs. \cite{roe17,sun20,kie21,sel21}.

As far as experimentally relevant observables are
concerned, the common consensus is that focusing on
few-body operators or sums thereof will be
sufficient.
As far as the generally 
out-of-equilibrium
initial states are concerned, the most common
examples are 
quantum quenches
\cite{dal16,gog16,mor18,ess16,vid16},
where an eigenstate 
(or an equilibrium ensemble)
of a so-called pre-quench
Hamiltonian serves as initial condition,
whose subsequent temporal evolution is
governed by some different, so-called 
post-quench Hamiltonian.

Within this general framework, 
and notwithstanding the above mentioned 
counter-examples, 
the presently prevailing expectation is that
thermalization after a quantum quench
can be taken for granted
if the post-quench system 
is non-integrable and obeys the ETH.
The main objective of the present paper is to
analytically show that this is not the case.

\section{Setup}
\label{s2}

For simplicity, we restrict ourselves to spin-models on a $d$-dimensional
hypercubic lattice $\Lambda:=\{1,...,L\}^d$ with a large but finite
number $| \Lambda |:=L^d$ of sites (degrees of freedom) 
and periodic boundary conditions
(some generalizations will be briefly mentioned later).
Furthermore, we mainly consider ``extensive'' 
(translation invariant) Hamiltonians
of the form 
\begin{eqnarray}
H=\sum_{i\in\Lambda} H_i \ ,
\label{1}
\end{eqnarray}
where the 
$H_i$ are translational copies of 
the same 
local
(few-body and short-range) 
operator
which only act nontrivially on lattice sites 
sufficiently close to
$i$.
Likewise, we often consider ``intensive''
observables of the form 
$A=| \Lambda |^{-1}\sum_{i\in\Lambda} A_i$.

A common example is the transverse-field 
Ising model (TFIM) with
\begin{eqnarray}
H_i = -g \, \sigma^x_i - \sum_{j\in\Lambda} J(|i-j|)\, \sigma^z_i\sigma^z_j 
\ ,
\label{2}
\end{eqnarray}
where $\sigma^x_i$ and $\sigma^z_i$ 
are Pauli matrices at lattice site $i$,
and $|i-j|$ is the natural distance 
between $i$ and $j$ on the (periodic) lattice. 
For instance, the interactions 
$J(|i-j|)$ may be unity if $i$ and $j$ 
are nearest neighbors, and zero otherwise.
A particularly interesting observable is
then the longitudinal magnetization
\begin{eqnarray}
M = \frac{1}{|\Lambda|}\sum_{i\in\Lambda} \sigma^z_i
\ .
\label{3}
\end{eqnarray}

\section{Illustration of the main result}
\label{s3}

Before working out the general theory, it may be instructive to
illustrate the main message of the paper by means of some particular
examples.

Therefore, let us focus on the following two-dimensional TFIM 
with periodic boundary conditions and nearest neighbor
interactions,
\begin{eqnarray}
H=-\sum_{\langle i j \rangle} \sigma_i^z\sigma_j^z - g \sum_{i\in\Lambda}
\sigma_i^x
\ ,
\label{n1}
\end{eqnarray}
where the first sum is over all nearest-neighbor sites 
on the two-dimensional (periodically closed)
square lattice $\Lambda$.
It is well-known that for $g$-values in the range $0<g<g_c\simeq 0.304$
this model is non-integrable and exhibits a phase transition
(spontaneous symmetry breaking) in the thermodynamic limit
$L\to\infty$  \cite{fra15}.
Put differently, in the canonical ensemble the order 
parameter (\ref{3}) vanishes above some critical 
temperature $T_c>0$ (which depends on $g$), and assumes 
a finite value below $T_c$.

Returning to arbitrarily large but finite system sizes $L$, 
let us now choose 
an eigenstate
of such a pre-quench 
Hamiltonian (\ref{n1}) (with $0<g<g_c$) as initial condition, 
which then evolves in time according to a post-quench 
Hamiltonian (on the same two-dimensional lattice $\Lambda$)
of the general from
\begin{eqnarray}
H_p  
& \!\! = \!\! &  
H_s + \lambda \sum_{i\in\Lambda} \sigma_i^z
\ ,
\label{n2}
\\
H_s 
& \!\! := \!\! & 
- \!\sum_{\langle i j \rangle}
\left(
J_x \sigma_i^x\sigma_j^x+J_y \sigma_i^y\sigma_j^y+J_z \sigma_i^y\sigma_j^y
\right)
- h \sum_{i\in\Lambda} \sigma_i^x
, \ \ \ \
\label{n3}
\end{eqnarray}
with largely arbitrary parameters $\lambda$, $J_{x,y,x}$, and $h$, 
except that $\lambda$ must be non-zero.
In particular, $H_s$
may be (but need not be) equal to the pre-quench Hamiltonian
(\ref{n1}), while the last term in (\ref{n2}) (with $\lambda\not=0$) is known to
exclude any phase transition for the post-quench system.
Moreover, many Hamiltonians of the general form (\ref{n2}), (\ref{n3}) 
are known or expected to be non-integrable and to obey the ETH
\cite{mor18,dal16,gog16,ess16,vid16,fra15}.

The main result of the present paper is the analytical prediction that 
in any such post-quench system there exists a large number 
of  pre-quench eigenstates which do {\em not} exhibit thermalization
after an initial quantum quench as described above.


\section{Pre-quench systems}
\label{s4}

On top of the assumptions around Eq. (\ref{1}), the pre-quench
Hamiltonian $H$ is required to exhibit a
so-called spin-flip (or $Z_2$) symmetry.

Let us first exemplify the general idea for the TFIM
in (\ref{1})-(\ref{3}):
One readily verifies that the Pauli matrices $\sigma_i^{a}$ 
are,  for any $a\in\{x,y,z\}$ and $i\in\Lambda$,
unitary operators, 
satisfying
$\sigma_i^x \sigma_i^z=-\sigma_i^z \sigma_i^x$.
It follows that 
also
$U := \prod_{i\in\Lambda} \sigma^x_i$ 
is a 
unitary operator, satisfying
$U \sigma_i^z =-\sigma_i^zU$ and
$U \sigma_i^x =\sigma_i^x U$ for 
any $i\in\Lambda$.
With (\ref{1})-(\ref{3}) this implies 
\begin{eqnarray}
UH & = & HU
\ ,
\label{4}
\\
UM & = & -MU
\ .
\label{5}
\end{eqnarray}


The extension to more general 
models 
in terms of
spin-operators $S_i^a$ 
is straightforward, well-known, and therefore
only briefly sketched:
Each summand contributing to $H$ now must 
consist of a product of factors of the form $S_i^a$
with $a\in\{x,y,z\}$ and $i\in \Lambda$,
so that the number of factors with the property 
$a\in\{y,z\}$ is even.
For instance, in the TFIM from (\ref{1}), (\ref{2})
these products are of the form $S^x_i$
and $S^z_i S^z_j$ with $S_i^a=\hbar \sigma_i^a/2$.
Analogously, $M$ may for the moment be any
operator with the following property:
Each summand contributing to $M$ must 
consist of a product of factors $S_i^a$
so that the number of factors with the property 
$a\in\{y,z\}$ is odd, as exemplified by (\ref{3}).
Defining $S^x:=\sum_{i\in\Lambda} S_i^x$ and
$U:=\exp\{i\pi S^x/\hbar\}$
implies again that $U$ is unitary,
and by exploiting the Hadamard lemma of
the Baker-Campbell-Hausdorff formula
one recovers again the key symmetries (\ref{4}) and (\ref{5}). 

Since $H$ and $U$ commute (see (\ref{4})), there exists a common
set of eigenvectors $|n\rangle$,
and since $U$ is unitary, all its eigenvalues are of unit modulus.
It follows that
$\langle n|U^\dagger MU|n\rangle=\langle n|M|n\rangle$,
while 
(\ref{5})
implies
$\langle n|U^\dagger MU|n\rangle=-\langle n|M|n\rangle$,
hence 
\begin{eqnarray}
\langle n|M|n\rangle = 0
\label{6}
\end{eqnarray} 
for all $|n\rangle$.
Likewise, $\langle n|H M|n\rangle$
can be identified with
$\langle n| U^\dagger H MU|n\rangle$,
implying with (\ref{4}) and (\ref{5}) that \cite{f0}
\begin{eqnarray}
\langle n|HM|n\rangle = 0
\label{7}
\end{eqnarray} 
for all pre-quench eigenstates $|n\rangle$.

Denoting by $E_n$ the 
corresponding 
eigenvalues of $H$ 
and by $\rhog:=Z^{-1} e^{-\beta H}$ 
with $Z:=\tr\{e^{-\beta H}\}$
the canonical ensemble, 
the expectation value $\tr\{\rhog M\}$ can be 
rewritten as
$\sum_n p_n \langle n|M|n\rangle$ with
$p_n:=Z^{-1} e^{-\beta E_n}$.
Eq. (\ref{6}) thus implies $\tr\{\rhog M\}=0$,
hence
the variance of $M$ in the canonical ensemble 
takes the form
\begin{eqnarray}
\sigma^2_{\! M\!,\mathrm{\beta},L}
:= \tr\{\rhog M^2\} = \sum_n p_n  \langle n|M^2|n\rangle
\ .
\label{8}
\end{eqnarray}

It is known (see below) or intuitively expected
that for many of the above 
specified spin-lattice models  there exists 
some suitable $M$,
for which the variance in (\ref{8})
converges in the thermodynamic limit to a non-vanishing value
$\sigma^2_{\! M\!,\mathrm{\beta}}:=\lim_{L\to\infty}\sigma^2_{\! M\!,\mathrm{\beta},L}$
below some critical temperature $\beta_c^{-1}>0$.
In this context, $M$ is then commonly
referred to
as order parameter, 
the variance in (\ref{8}) 
as its thermal fluctuations,
and their finite value for 
large $L$ as 
long-range order,
announcing
a phase transition
via spontaneous symmetry breaking (SSB)
in the thermodynamic limit.
Here, we 
adopt these 
standard
notions without
recapitulating the well-known 
underlying physics,
since we are actually 
only interested in
the fact that 
$\sigma^2_{\! M\!,\mathrm{\beta},L}\simeq \sigma^2_{\! M\!,\mathrm{\beta}}>0$
for
sufficiently large (but finite) $L$
if $\beta>\beta_c$.

Indeed, such a behavior of 
$\sigma^2_{\! M\!,\mathrm{\beta},L}$
has been rigorously 
derived
for a considerable variety of short-range spin-lattice models 
in $d\geq 2$ dimensions,
quite often with an order parameter $M$ which is identical or 
similar to the magnetization in (\ref{3}),
see e.g. Refs. \cite{dys69,dys78,fro78a,fro78b,fro78c,ken85,dat96,bor96} 
and further references therein.
Incidentally, these rigorous results also cover various
generalizations of the model class specified around 
(\ref{1}),
including natural instead of periodic boundary conditions,
other lattice geometries,
one-dimensional models with long-range interactions,
and lattice-gas instead of spin models.

Eq. (\ref{8}) implies that
$\langle n|M^2|n\rangle\geq \sigma^2_{\! M\!,\mathrm{\beta}}/2$ 
for at least {\em one} eigenstate $|n\rangle$, given 
$L$ is sufficiently large and $\beta>\beta_c$.
Focusing on $\beta\to\infty$, this pertains,
in particular, to the ground state of $H$
(or at least to one of them in case of degeneracy).
More generally, 
it seems reasonable to expect
that there actually exist {\em many} eigenstates 
$|n\rangle$ for which $\langle n|M^2|n\rangle$
exceeds some suitable threshold, for instance
$\sigma^2_{\! M\!,\mathrm{\beta}}/2$.
The straightforward but somewhat tedious 
verification of this expectation is worked out in
the Appendix, showing that the number of
those eigenstates is actually exponentially large in $L$.
Numerically, this issue is at (or even beyond)
current feasibility limits \cite{fra15,mon16,fra16}.
In particular, the sophisticated numerical explorations 
in Ref. \cite{fra16}  suggest that our above 
expectation may actually apply even to
{\em all} eigenstates $|n\rangle$ with sufficiently 
low energies, e.g., $E_n<\tr\{\rhog H\}$ 
for some $\beta>\beta_c$.
Analytically, the same conclusion can also be recovered
under the assumption that the diagonal matrix elements 
$\langle n|M^2|n\rangle$ obey the ETH \cite{fra16}.

\section{Post-quench systems}
\label{s5}

As announced in the introduction, we focus on the most
common quantum quench scenario,
where the initial condition $|\psi(0)\rangle$ 
is given by an eigenstate $|n\rangle$ of the 
pre-quench Hamiltonian $H$, 
while the actual dynamics $|\psi(t)\rangle=e^{-iH_pt/\hbar}|n\rangle$
is governed by a different, post-quench 
Hamiltonian $H_p$ \cite{dal16,gog16,mor18,ess16,vid16}.
In particular, we are {\em not} interested in the 
thermalization 
of
the pre-quench system.
Moreover, we restrict ourselves to initial
states $|n\rangle$ for which 
$\langle n|M^2|n\rangle$
exceeds some $L$-independent threshold
value $\mu$, for instance 
$\mu=\sigma^2_{\! M\!,\mathrm{\beta}}/2$
(see above).
Finally, instead of a pre-quench eigenstate
we will also consider initial conditions in the 
form of a pre-quench equilibrium ensemble.

As exemplified by (\ref{n2}) and (\ref{n3}), 
the post-quench Hamiltonian is required 
to be of the form
\begin{eqnarray}
H_p = H_s + \lambda V
\ ,
\label{9}
\end{eqnarray}
where $H_s$ is of the same general structure as the
Hamiltonians discussed around Eqs. (\ref{1}) and 
(\ref{4}), and thus obeys again symmetry relations 
(hence the index ``$s$'') analogous 
to (\ref{4}) and (\ref{7}), i.e. \cite{f0},
\begin{eqnarray}
\langle n|H_s M|n\rangle = 0
\ .
\label{10}
\end{eqnarray}
Furthermore, $\lambda$ must
be non-zero and $L$-independent, while
$V$ is required to be of the form 
\begin{eqnarray}
V=|\Lambda| M
\ ,
\label{11}
\end{eqnarray}
where $M$ is the order parameter of the 
pre-quench system $H$.
Hence, $V$ is usually an extensive quantity, 
as for instance in (\ref{3}) and (\ref{n2}).

In general, $H_s$ may but need not agree with
the pre-quench Hamiltonian $H$.
In particular, and as exempliefied by (\ref{n2}) and (\ref{n3}), $H_s$ may be 
a Heisenberg-type model \cite{f1},
possibly with some anisotropy (e.g. XXZ- or XY-models) 
and/or next-nearest-neighbor interactions etc.
Moreover, $H_s$ may for instance contain
-- similarly as in 
(\ref{n3}) --
a contribution proportional to $\sum_{i\in \Lambda} S^x_i$.
In fact, since only the property (\ref{10}) will
actually be needed below, the post-quench
system may 
even exhibit some disorder, possibly giving rise to MBL.

Importantly, the perturbation $V$ in (\ref{9}) breaks 
the symmetry 
of the unperturbed 
$H_s$. 
Hence, the post-quench system $H_p$ generically does not exhibit 
long-range order, and no SSB and phase transitions
will occur in the thermodynamic limit.

Furthermore, for many of the above specified 
examples, the Hamiltonian $H_p$ 
is
commonly
expected to be non-integrable and to obey the ETH
\cite{mor18,dal16,gog16,ess16,vid16,fra15,mon16,fra16},
though rigorous proofs
are usually not available, 
and even the precise meaning of ``integrability'' 
is still not entirely clear \cite{dal16,gog16}.
Since these issues are not at the focus of our present 
paper, we tacitly take for grated those commonly expected properties.

\section{Non-thermalization}
\label{s6}

In a first step, the essential arguments are worked out
in case the initial condition is given by an eigenstate
$|n\rangle$ of the pre-quench Hamiltonian.
Then, the modifications for initial conditions in the form of
a canonical or microcanonical 
pre-quench equilibrium ensemble are addressed.
Finally, additional details and physical arguments are provided,
for simplicity focusing again on pre-quench eigenstates.

\subsection{Pre-quench eigenstates}
\label{s61}

Eqs. (\ref{6}), (\ref{9}), and (\ref{11}) imply
\begin{eqnarray}
\langle n|H_p|n\rangle 
=
\langle n|H_s|n\rangle
\ .
\label{12}
\end{eqnarray}
Furthermore, we can conclude from Eqs. (\ref{9})-(\ref{11}) that
\begin{eqnarray}
\langle n|H_p^2|n\rangle 
& = &
\langle n|H_s^2|n\rangle
+
\lambda^2 \langle n|V^2|n\rangle
\ .
\label{13}
\end{eqnarray}
Combining (\ref{12}) and (\ref{13}) yields
\begin{eqnarray}
\sigma^2_{\! p,n} & = & \sigma^2_{\! s,n} + \lambda^2  \langle n|V^2|n\rangle
\ ,
\label{14}
\\ 
\sigma^2_{\! p,n} & := &\langle n|H_p^2|n\rangle - \langle n|H_p|n\rangle^2
\ ,
\label{15}
\\
\sigma^2_{\!s,n} & := & \langle n|H_s^2|n\rangle - \langle n|H_s|n\rangle^2
\ .
\label{16}
\end{eqnarray}
Observing (\ref{11}) and $\sigma^2_{\! s,n}\geq 0$, we thus obtain
\begin{eqnarray}
\sigma^2_{\! p,n} \geq  
|\Lambda|^2 \lambda^2  \langle n|M^2|n\rangle
\ .
\label{17}
\end{eqnarray}
As detailed above (\ref{9}), 
$\langle n|M^2|n\rangle$
is lower bounded by an $L$-independent 
constant on the order of  
$\sigma^2_{\! M\!,\mathrm{\beta}}$,
yielding
\begin{eqnarray}
\sigma_{\! p,n} \geq  
c\, |\Lambda|
\ ,
\label{18}
\end{eqnarray}
where $c$ is on the order of 
$|\lambda| \sigma^2_{\! M\!,\mathrm{\beta}}$
and independent of the system size $|\Lambda|$.

A well-established prerequisite for thermalization
is that the system's energy distribution 
must be sufficiently
narrow, 
i.e., the energy spread must be 
small (subextensive) in comparison with
the typical (extensive) system energies themselves
\cite{dal16,gog16,rei08,sre96,sre99,rig08,bri10,ess16}.
On the other hand, Eqs. (\ref{15}) and (\ref{18})
tell us that the energy spread 
(standard deviation) $\sigma_{\! p,n}$ is (at least) 
extensive in the system size $|\Lambda|$.
As a consequence, the system cannot 
exhibit thermalization.

Before expounding in more detail this very condensed 
line of reasoning, we next address some modifications of the 
considered initial states.

\subsection{Canonical and microcanonical pre-quench ensembles}
\label{s62}

Instead of a pre-quench eigenstate, let us now turn to 
initial conditions in the form of a canonical ensemble. 
As detailed above Eq. (\ref{8}), we are thus dealing
with a mixed initial state (density operator) of the form
\begin{eqnarray}
\rho(0) = \rhog = \sum_{n=1}^N p_n \, |n\rangle\langle n|
\label{n4}
\end{eqnarray}
with $p_n:=Z^{-1}e^{-\beta E_n}$.
Similarly as in (\ref{12}),  
the post-quench energy expectation value
\begin{eqnarray}
\langle H_p\rangle:=\tr\{\rhog H_p\}
\label{n5}
\end{eqnarray}
thus satisfies the relations
\begin{eqnarray}
\langle H_p\rangle
=
\sum_{n=1}^N p_n \langle n|H_p|n\rangle
= 
\sum_{n=1}^N p_n \langle n|H_s|n\rangle 
=
\langle H_s\rangle
\ .
\label{n6}
\end{eqnarray}
Likewise, the corresponding second moment 
now takes, similarly in (\ref{13}), the form
\begin{eqnarray}
\langle H^2_p\rangle
= 
\langle H^2_s\rangle + \lambda^2 \langle V^2\rangle
\ .
\label{n7}
\end{eqnarray}
Concerning the post-quench energy variance
\begin{eqnarray}
\sigma^2_{\! p} & := &\langle H_p^2\rangle - \langle H_p\rangle^2
\label{n8}
\end{eqnarray}
we then can conclude, similarly as in (\ref{14})-(\ref{17}) that
\begin{eqnarray}
\sigma^2_{\! p} \geq  
|\Lambda|^2 \lambda^2  \langle M^2\rangle
\label{n9}
\end{eqnarray}
and with (\ref{8}) that
\begin{eqnarray}
\sigma^2_{\! p} \geq  
|\Lambda|^2 \lambda^2  \sigma^2_{\! M\!,\mathrm{\beta},L}
\ .
\label{n10}
\end{eqnarray}
If the temperature $\beta^{-1}$ of the canonical ensemble 
in (\ref{n4}) is smaller than
the critical temperature $\beta_c^{-1}$ (see below (\ref{8})),
thermalization can thus again be ruled out analogously
as below (\ref{18}).

Closely related to the discussion at the end of Sec.~\ref{s4}
we  remark that the usually expected equivalence 
of ensembles has not been rigorously proven until now
for systems which exhibit long-range order
\cite{kuw20a,tas18,tou15}. 
Yet it seems reasonable to expect that, 
at least qualitatively, the mere existence 
of long-range order would also 
be recovered in the microcanonical 
ensemble in cases where it provably
occurs in the canonical ensemble  
\cite{dys69,dys78,fro78a,fro78b,fro78c,ken85,dat96,bor96}.
If so, one can readily modify the above line of reasoning
to shown non-thermalization also in cases where the initial condition
$\rho(0)$ is given by a microcanonical ensemble
with a concomitant temperature which is smaller than 
$\beta_c^{-1}$.

\subsection{More detailed discussion}
\label{s63}

A more detailed version of the
arguments at the end of Sec. \ref{s61}
is as follows:
Denoting the eigenvalues and eigenvectors of
$H_p$ by $\tilde E_n$ and $|\tilde n\rangle$,
a projective measurement of the observable
$H_p$ yields the outcome 
$\tilde E_n$ with probability 
$\tilde p_n:=|\langle \tilde n|\psi\rangle|^2$
for any given (normalized) system state $|\psi\rangle$.
On the average over many repetitions of the 
measurement, the mean value 
is thus
$\sum_n \tilde p_n \tilde E_n=\langle\psi |H_p|\psi\rangle$ and
the second moment 
$\sum_n \tilde p_n \tilde E_n^2=\langle\psi |H_p^2|\psi\rangle$.
Hence, $\sigma^2_{\! p,n}$ in (\ref{15}) is the variance 
when repeatedly measuring the (post-quench) system 
energy $H_p$ in the initial state $|\psi(0)\rangle=|n\rangle$ 
of the post-quench dynamics.
But since the energy is a conserved quantity, 
$\sigma^2_{\! p,n}$ is at the same time the 
energy variance for any later
system state $|\psi(t)\rangle$.

Accordingly, the energy spread (standard deviation) 
is given by $\sigma_{\! p,n}$ for all times
$t$ and thus also in the long-time limit,
and scales according to (\ref{18}) at least linearly 
with the system size $|\Lambda|$.

As said in the introduction, a necessary condition for 
thermalization is that the  long-time behavior of 
all relevant (measurable) observables must be well-approximated by
the corresponding microcanonical expectation 
values.
One such relevant observable is the system energy
$H_p$ (or its ``intensive'' counterpart in 
Eq. (\ref{21}) below).
Furthermore, for our system states $|\psi(t)\rangle$
with initial condition $|n\rangle$, the pertinent 
microcanonical ensemble $\rhomic$ must
satisfy the usual condition
\begin{eqnarray}
E:=\tr\{\rhomic H_p\} = \langle n|H_p|n\rangle
\ ,
\label{19}
\end{eqnarray}
i.e., the ``microcanonical energy window'' must be
chosen so that the energy of the actual system under
consideration is correctly reproduced.
Note that in spite of the fact that $|n\rangle$ may be
the ground state of the pre-quench Hamiltonian $H$, 
the corresponding post-quench energy (\ref{19}) is 
in general {\em not} close to the ground state of the 
post-quench Hamiltonian $H_p$.

Though the post-quench system is in general not
assumed to be at thermal equilibrium, 
we may nevertheless ask how it 
would behave 
at thermal equilibrium.
Then, the post-quench system would, essentially by 
definition, comply with the textbook microcanonical 
formalism.
In particular, the microcanonical energy variance
\begin{eqnarray}
\sigma^2_{\!\mic}:=
\tr\{\rhomic H^2_p\} - E^2
\label{20}
\end{eqnarray}
is then known to be -- also depending on the actual choice 
of the  microcanonical energy window \cite{tas18} -- 
at most on the order of $[\kB T(E)]^2$,
where $\kB$ is Boltzmann's constant and $T(E)$ the 
microcanonical temperature corresponding to the
system energy $E$ from (\ref{19}).
Under any reasonable ``upscaling procedure'',
the temperature $T(E)$
is furthermore expected to approach some finite value
in the thermodynamic limit $|\Lambda|\to\infty$
(or at least to grow slower than $|\Lambda|$).
Hence, the energy spread of the actual system
is, according to (\ref{18}), incompatible with the
corresponding spread 
$\sigma_{\!\mic}$ at thermal equilibrium
\cite{f2}.

Again from a different viewpoint, let us 
consider the ``intensive'' observable 
(energy density)
\begin{eqnarray}
A:=H_p/|\Lambda|
\ ,
\label{21}
\end{eqnarray}
see also below (\ref{1}).
Given $A$ is a relevant (experimentally measurable)
observable, the same must apply to $A^2$.
(One simply has to square the outcome of 
each (projective) measurement of $A$
\cite{fx}.)
Exploiting (\ref{15}), (\ref{18}), and (\ref{19})-(\ref{21}),
one readily verifies that
\begin{eqnarray}
\langle n|A^2|n\rangle - \tr\{\rhomic A^2\}
= \frac{\sigma^2_{\! p,n} - \sigma^2_{\!\mic}}{|\Lambda|^2}
\geq c^2 - \frac{\sigma^2_{\!\mic}}{|\Lambda|^2}
\ .
\label{22}
\end{eqnarray}
As before, the right hand side 
generically approaches
a positive value
for sufficiently large systems,
hence the
difference on the left hand side is 
not negligible compared to the the two 
expectation values themselves.
In conclusion, the system does 
not exhibit thermalization.


As far as the non-thermalization of other observables 
than the energy density in (\ref{21}) is concerned, obtaining rigorous 
statements turns out
to be rather difficult, while non-rigorous
arguments are straightforward and still quite convincing:
For simplicity, let us focus on observables $A$ which satisfy 
the ETH (non-thermalization in the absence of ETH is 
quite common anyway).
Then, the long-time average of $\langle \psi(t)|A|\psi(t)\rangle$
is well-know \cite{dal16,sre96,sre99,fra16}
to be closely approximated by
$\sum_m \tilde p_m\, {\cal A}(\tilde E_m)$,
where $\tilde p_m:=|\langle\tilde m| n\rangle|^2$ 
are the populations of the post-quench 
energy levels $|\tilde m\rangle$ by the initial state 
$|\psi(0)\rangle=|n\rangle$, $\tilde E_m$ are the 
corresponding post-quench eigenvalues, 
and ${\cal A}(x)$ is a smooth function of its argument $x$.
Likewise, the microcanonical expectation value 
$\tr\{\rhomic A\}$ is closely approximated 
by 
${\cal A}(E)$, where $E$ is given by (\ref{19}).
A necessary condition for thermalization is that
the long-time average must be close to the
microcanonical expectation value, i.e.,
the approximation
\begin{eqnarray}
\sum_m \tilde p_m\, {\cal A}(\tilde E_m) = {\cal A}(E)
\label{23}
\end{eqnarray}
must be fulfilled very well.
However, since the energy distribution of the initial state 
$|n\rangle$ is {\em not\,} narrow according to (\ref{15}) 
and (\ref{18}), the deviations of ${\cal A}(\tilde E_m)$
from ${\cal A}(E)$ are in general {\em not\,} negligible 
on the left hand side of (\ref{23}), and hence 
the approximation will not be fulfilled 
sufficiently well.

In hindsight it may be worthwhile to recall
that if a system thermalizes
it is understood that every relevant (measurable) 
quantity must exhibit thermalization.
In turn, in order to show non-thermalization it is 
sufficient that at least one measurable
quantity exhibits non-thermalization.
As argued above, the energy spread is 
such a quantity, hence our demonstration
of non-thermalization is completed.
Nevertheless, it is interesting to consider
the behavior of other quantities, 
as was done in the previous paragraph.
Though we were not able to show
their non-thermalization with similar rigor 
as for the energy spread, this does not
undermine  \cite{f3} our already completed
(rigorous) demonstration of 
non-thermalization.

\section{Conclusions and comparison with related works}
\label{s7}

A large class of quite common spin-lattice models 
have been identified which do not exhibit 
thermalization after a quantum quench.
Among them are many examples which are generally
considered to be non-integrable and to obey the ETH.
The salient point is that the pre-quench system
must exhibit long-range order in the canonical ensemble
(due to an underlying spin-flip symmetry),
hence 
an analogous
property is inherited by exponentially many
pre-quench eigenstates (including the ground state),
and (presumably) also by the corresponding microcanonical
ensemble.
Choosing one of these eigenstates or equilibrium 
ensembles as initial condition, and properly
incorporating the concomitant order parameter into
the post-quench Hamiltonian, then gives rise to a non-narrow
post-quench energy distribution, which is known to be
incompatible with the assumption that the system 
exhibits thermalization in the long run
\cite{dal16,gog16,rei08,sre96,sre99,rig08,bri10,ess16}.

The general setup and reasoning in the present paper
are somewhat similar to those in Ref. \cite{alb17},
but there are also significant differences.
Most importantly, the main focus of Ref. \cite{alb17} 
is on the implications of long-range order for 
so-called prethermalization phenomena, 
not on the possible absence of thermalization 
in the ``true'' long-time limit.
Moreover, the present paper addresses systems
which exhibit long-range order at non-vanishing 
temperatures, while Ref. \cite{alb17} mainly considers
systems where long-range order is 
present only at zero temperature (corresponding to a
quantum phase transition in the thermodynamic limit).
Finally, some of the arguments adopted in 
Ref. \cite{alb17} are less rigorous than in the
present paper.

Note that the requirement of a narrow energy 
distribution in the context of thermalization is 
well-established.
However, it would not be right to say \cite{f3} that,
as a consequence, the requirement must 
always be fulfilled,
nor that the requirement already implies as 
a more or less obvious consequence the main 
result of our paper, namely the existence of 
relevant examples with a non-narrow energy 
distribution.

Overall, our so-far understanding of thermalization 
is predominantly based on numerical evidence
and non-rigorous arguments 
\cite{mor18,dal16,gog16,lan16,ued20,nan15}.
The present main result belongs to the still 
rather scarce 
analytical statements
in this context.
The key ingredients for its derivation were 
symmetry considerations and previously established 
proofs of canonical long-range order for instance in Refs. 
\cite{dys69,dys78,fro78a,fro78b,fro78c,ken85,dat96,bor96}.

Incidentally, a general argument that the energy
distribution must be narrow after a quantum quench
has been provided in the Supplemental Information
of Ref. \cite{rig08}, however explicitly assuming
``the absence of long-range correlations''.
A similar argument in Appendix A of Ref. \cite{ess16}
assumes a closely-related, so-called 
cluster decomposition property.
Apparently, both assumptions are 
not fulfilled
by our examples.
In this sense, the present paper is complementary 
to Refs. \cite{rig08,ess16} and  numerous subsequent 
works which built on them.

\vspace*{1cm}
\begin{acknowledgments}
Invaluable discussions with Lennart Dabelow,
J\"urgen Schnack, Fabian Essler, and Michael Kastner 
are gratefully acknowledged.
This work was supported by the 
Deutsche Forschungsgemeinschaft (DFG)
within the Research Unit FOR 2692
under Grant No.~355031190,
and by the International Centre for Theoretical Sciences 
(ICTS) during a visit for the program -  
Thermalization, Many-body localization and Hydrodynamics 
(Code: ICTS/hydrodynamics2019/11).
\end{acknowledgments}

\appendix
\section{Eigenstates with non-negligible variance of $M$}
\label{app1}

Taking for granted that the order parameter $M$ is an 
intensive observable, as exemplified by (\ref{3})
(see also below (\ref{8})), there exists some finite
upper bound of the operator norm $\norm{M^2}$
for all $L$. For instance, (\ref{3}) 
readily implies that $\norm{M^2}$ is upper bounded
by $\Lambda^{-2}\sum_{ij\in\Lambda}\norm{\sigma_i^z\sigma_j^z}$
and with $\norm{\sigma_i^z\sigma_j^z}\leq \norm{\sigma_i^z}\norm{\sigma_j^z}=1$
that $\norm{M^2}\leq 1$ for all $L$. Without loss of generality
we denote by $a$ the smallest such bound of $\norm{M^2}$,
implying
\begin{eqnarray}
\langle n |M^2|n\rangle\leq a
\label{a1}
\end{eqnarray}
for all $n$ and all $L$.

Since $\sigma^2_{\! M\!,\mathrm{\beta},L}$ converges to 
$\sigma^2_{\! M\!,\mathrm{\beta}}$ for large $L$ 
and any given $\beta>\beta_c$ (see below (\ref{8})), 
we know that
\begin{eqnarray}
\sigma^2_{\! M\!,\mathrm{\beta},L}\geq  (3/4)\, \sigma^2_{\! M\!,\mathrm{\beta}}  
\label{a2}
\end{eqnarray}
for all sufficiently large $L$.

Finally, for any given $L$ and $\beta>\beta_c$, we denote 
by $I_{\!\beta\!,L}$ the set of all indices $n$ with the 
property that
\begin{eqnarray}
\langle n |M^2|n\rangle\geq \sigma^2_{\! M\!,\mathrm{\beta}}/2
\label{a3}
\end{eqnarray}
and by $N_{\!\beta\!,L}$  the number of elements
contained in the set $I_{\!\beta\!,L}$.
In other words, $N_{\!\beta\!,L}$ represents the number
of eigenstates $| n\rangle$ with the property (\ref{a3}).

From now on we tacitly restrict ourselves to sufficiently large
$L$, so that both (\ref{a1}) and (\ref{a2}) are fulfilled,
and to temperatures $\beta^{-1}$ below the critical 
temperature $\beta_c^{-1}$, so that 
$\sigma^2_{\! M\!,\mathrm{\beta}} >0$ (see below (\ref{8})).

Without loss of generality we furthermore assume that the 
eigenvalues $E_n$ are ordered by magnitude 
and that the indices $n$ run from $1$ 
to some upper limit $N$
(which is exponentially large in the system 
size $|\Lambda|=L^d$ for any given spin-model, but
which in principle may also be infinite).
Accordingly, (\ref{8}) implies
\begin{eqnarray}
\sigma^2_{\! M\!,\mathrm{\beta},L}
& = & 
\sum_{n=1}^N p_n  \langle n|M^2|n\rangle = K_1+K_2
\ ,
\label{a4}
\\
K_1 & := & \sum_{n\in I_{\!\beta\!,L}} p_n  \langle n|M^2|n\rangle
\ ,
\label{a5}
\\
K_2 & := & \sum_{n\not\in I_{\!\beta\!,L}} p_n  \langle n|M^2|n\rangle
\ ,
\\
p_n & := & \frac{e^{-\beta E_n}}{\sum_{m=1}^N e^{-\beta E_m}}
\ .
\label{a7}
\end{eqnarray}

With (\ref{a1}) it follows that $K_1$ in (\ref{a5}) is upper bounded by 
$\sum_{n\in I_{\!\beta\!,L}} p_n a$, and since $E_1\leq E_n$ 
and thus $p_n\leq p_1$ for all
$n$ that
\begin{eqnarray}
K_1 \leq \sum_{n\in I_{\!\beta\!,L}} p_1 a  = 
p_1 a\, N_{\!\beta\!,L}
\ .
\label{a8}
\end{eqnarray}
On the other hand, since all $n\not\in I_{\!\beta\!,L}$ satisfy 
$\langle n |M^2|n\rangle\leq b_\beta := \sigma^2_{\! M\!,\mathrm{\beta}}/2$ according to
(\ref{a3}), we can conclude that
\begin{eqnarray}
K_2 \leq \sum_{n\not\in I_{\!\beta\!,L}} p_n  b_\beta
\leq \sum_{n=1}^N p_n  b_\beta = b_\beta =\sigma^2_{\! M\!,\mathrm{\beta}}/2
\ .
\label{a9}
\end{eqnarray}
With (\ref{a4}), (\ref{a8}), (\ref{a9}) it follows that
\begin{eqnarray}
p_1 a\, N_{\!\beta\!,L}\geq K_1= \sigma^2_{\! M\!,\mathrm{\beta},L} - K_2 \geq
 \sigma^2_{\! M\!,\mathrm{\beta},L}
-  \sigma^2_{\! M\!,\mathrm{\beta}}/2
\ \ \ 
\label{a10}
\end{eqnarray}
and with (\ref{a2}) and (\ref{a7}) that
\begin{eqnarray}
N_{\!\beta\!,L}
& \geq &  
Z\, \frac{\sigma^2_{\! M\!,\mathrm{\beta}}}{4\, a}
\ ,
\label{a11}
\\
Z & := & \sum_{m=1}^N\e^{-\beta(E_m-E_1)}
\ .
\label{a12}
\end{eqnarray}
Disregarding extremely small temperatures $\beta^{-1}$, 
the number of energies $E_n$ with the property that 
$\beta(E_n-E_1)\approx 1$ may be
expected to grow exponentially with the system size $L$, 
hence $Z$ will be exponentially large in $L$ as well.
Alternatively, $Z$ in (\ref{a12}) may be viewed as a canonical
partition function with the natural convention that 
energies are measured relatively to the ground state energy 
$E_1$. 
Taking for granted generic thermodynamic properties, 
it follows that
$Z=e^{-\beta F}\!$, where $F$ is the free energy, and where
$\beta F$ is 
an extensive, negative quantity, implying once again
that $Z$ is exponentially large in the system size $L$.

On the other hand, the last factor on the right hand side of
(\ref{a11}) is independent of $L$. Moreover, this quantity is
dimensionless, positive, and generically not expected to
be extremely small compared to unity (except if
the temperature $\beta^{-1}$ is very close to the
critical temperature $\beta_c^{-1}$).

Altogether,
the number of eigenstates $|n\rangle$, which 
exhibit a non-negligible variance of $M$ in the sense of
(\ref{a3}), is given by the left hand side of (\ref{a11}) 
and is found to grow exponentially with $L$ 
for any given temperature  $\beta^{-1}$ below
(and not too close to) $\beta_c^{-1}$.



\begin{thebibliography}{11}
\bibitem{f0}
In fact, the derivation of (\ref{6}) and (\ref{7}) only
requires that $|n\rangle$ is an eigenvector of $U$,
not necessarily of $H$, and likewise for $H_s$ in
(\ref{10}).
\bibitem{f1}
For one-dimensional models,
the pre-quench Hamiltonian $H$ must 
exhibit long-range (e.g. algebraically decaying)
interactions, as mentioned below Eq. (\ref{8})
and exemplified in Ref. \cite{fra16}.
\bibitem{f2}
When employing a canonical
instead of a microcanonical ensemble,
the energy spread (standard deviation) 
is known to scale as $|\Lambda|^{1/2}$,
which is again incompatible with (\ref{18}).
\bibitem{fx}
The situation is remarkably similar to
the proposition in Refs.  \cite{sre96,sre99,fra16}
that if an observable $A$ obeys the ETH, 
the same applies to $A^2$.
\end{thebibliography}
\end{document}